\documentclass[aps,prd,twocolumn]{revtex4-1}

\usepackage{CJKutf8}
\usepackage{graphicx}
\usepackage{url}
\usepackage{siunitx}

\begin{document}
    \begin{CJK*}{UTF8}{gbsn}
	\title{Blazar -- IceCube neutrino association revisited}
	\date{Last updated XXX; in original form YYY}
	\author{Jia-Wei Luo (罗佳伟)}
	\author{Bing Zhang (张冰)}
	\affiliation{Department of Physics and Astronomy, University of Nevada Las Vegas, NV 89154, USA}
	\begin{abstract}
		The reported association of high-energy neutrino event IceCube-170922A and blazar TXS 0506+056 has sparked discussion about blazars as sources of cosmic neutrinos. In this paper, we use publicly released IceCube data and blazar locations from Roma-BzCat to test spatial correlation between neutrino events and blazar locations. We also scrutinize the correlation between the $\gamma$-ray flux and neutrino flux of blazars by applying a temporal filter onto the data based on the Fermi monitored source list. We find no compelling evidence to prove blazars as the main source of cosmic neutrinos, as known before the detected IceCube-170922A / TXS 0506+056 association. While we do not rule out the association between IceCube-170922A and TXS 0506+056, the significance level we obtained is not high enough to claim a strong association. If such an association is real, a special physical condition is desired to allow a small fraction of blazars to become bright neutrino sources.
	\end{abstract}
	\maketitle
	\end{CJK*}
	\section{Introduction}
	\label{sec:introduction}

	Since the discovery of the diffuse flux of high-energy cosmic neutrinos by the IceCube Collaboration in 2013 \cite{icecube2013evidence}, many efforts have been made to identify the source of this diffuse neutrino flux. The distribution of the arriving directions of high-energy cosmic neutrinos does not deviate much from being isotropic \cite{aartsen2017all}. Before the detection of a high-energy neutrino event at the direction of blazar TXS 0506+056 in coincidence with a $\gamma$-ray flare of the same blazar \cite{icecube2018multimessenger}, 
	%and the recent discovery of possible association between FSRQ PKS 1502+106 and IceCube-190730A \cite{lipunov2019icecube}. and 
	no compelling evidence showed any association between cosmic neutrinos and any type of sources \cite{aartsen2017all}.
		
	Neutrinos are thought to be created alongside with $\gamma$-ray photons in physical processes \cite{2018PhT....71j..36M}. As a result, blazars, along with many other $\gamma$-ray sources, including $\gamma$-ray bursts, supernovae, star-burst galaxies, have long been proposed as possible sources of cosmic neutrinos \cite{albrecht2017neutrino, murase2019high}. While a significant amount of research work has been done by analyzing available IceCube data \cite{aartsen2017all,aartsen2017contribution,krauss2018fermi,hooper2019active}, the lack of spatial and temporal coincidences of certain types of sources (e.g. $\gamma$-ray bursts) and high-energy cosmic neutrino events has led to ruling out these types $\gamma$-ray sources as the dominant sources of the observed diffuse high-energy cosmic neutrino flux.
	
	A blazar TXS 0506+056 was claimed to be associated with the IceCube-170922A high-energy neutrino event while it was in a $\gamma$-ray flaring state \cite{icecube2018multimessenger}. An archival search revealed further association of the blazar with neutrino events, albeit at lower energies \cite{icecube2018neutrino}. This makes blazars an attractive candidate source for high-energy cosmic neutrinos detected by IceCube. However, there are many $\gamma$-ray sources brighter than TXS 0506+056, yet none of them are found to be associated with neutrinos. TXS 0506+056 is only the 183rd-brightest $\gamma$-ray source in the Fermi 4FGL catalog \cite{2019arXiv190210045T}.
	
	There are studies using IceCube data only to find neutrino point sources \cite{turley2018coincidence,krauss2018fermi,hooper2019active}, but no association between neutrino and known sources have been found, including TXS 0506+056. Indeed, the neutrino events clustering in the direction of TXS 0506+056 give significance of only $2.9\sigma$. It is the $\gamma$-ray flare detection at the same time that makes the association more credible to reach $3.7\sigma$ \cite{icecube2018multimessenger}. Theoretical modeling also suggested that it is not straightforward to interpret IceCube-170922A as emission from TXS 0506+056 \cite{gao2019modelling,halzen2019neutrino,padovani2018dissecting,sahakyan2018lepto,murase2018blazar,liu19}, and it is essentially impossible to account for the 2015 lower-energy flare \cite{petropoulou2019neutrinos}.

	In this paper, we revisit the blazar-neutrino problem by making use of the IceCube data, the Fermi LAT data, and the Roma-BzCat blazar catalog to investigate possible neutrino-blazar associations. We employ two approaches to test the association between blazar and neutrino event locations. The first approach is to match blazars in the directions of neutrino events, and the second approach is to match neutrino events in the directions of blazars. 
	
	\section{Coincidence search of blazars in neutrino event directions}
	\label{sec:Blazarinneutrino}
	
	In this section, we use all blazar directions from the fifth edition of the blazar catalog ROMA-BzCat \cite{roma5th} and the IceCube catalog of alert events up through IceCube-170922A at
    https:// icecube.wisc.edu/science/data/TXS0506\_alerts \cite{Icecube2018alerts}. We adopt alert events, since we believe that these events better represent IceCube-170922A that was claimed to be associated with TXS 0506+056. A total of 3561 blazars and 45 high-energy neutrino alerts are included.
	
\begin{figure}[ht]
		\centering
		\includegraphics[width=0.5\textwidth]{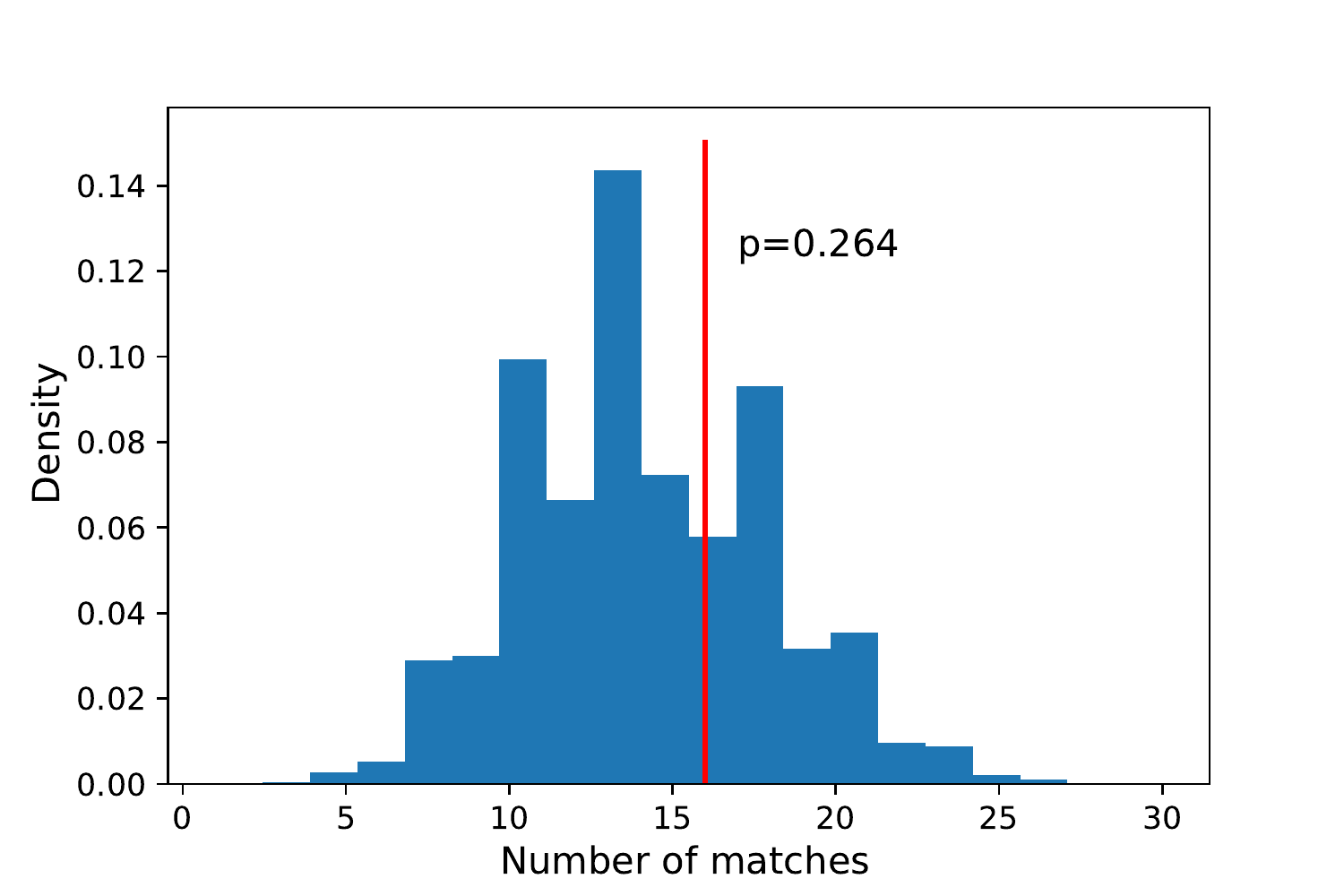}
		\caption{The distribution of the simulated number of matches between the directions of blazars in ROMA-BzCat and the directions of IceCube neutrino alert events. The vertical line denotes the observed value of $16$.}
		\label{fig:match}
	\end{figure}

    We match the directions of the blazars with the directions of the neutrino alerts to count the total number of blazars which lie within the error region of each neutrino event. There are $16$ matches. We then randomly generate the same amount of blazar directions as in the blazar catalog and conduct the same matching search. A total of 10 000 trials are carried out, and we compare the simulated numbers of matches with the observed one. As shown in Fig. \ref{fig:match}, the observed number of matches does not deviate much from the average value of the simulated ones; thus, an association between blazars and neutrino alert directions is not supported by this method. The blazar catalog has a deficit of sources along the Galactic plane (due to an observational selection effect). However, physically blazars should be distributed isotropically in the Universe. In our simulation, if we intentionally drop out the simulated blazars near the Galactic plane, the total number of matches slightly drops, but the probability of matches essentially remains the same.
	
	We also scrutinize the coincidence of IceCube-170922A with blazar TXS 0506+056 by comparing the minimum angular distance from the simulated blazars to IceCube-170922A in each simulation with the actual angular distance between TXS 0506+056 and IceCube-170922A.
	
\begin{figure}[ht]
	\includegraphics[width=0.49\textwidth]{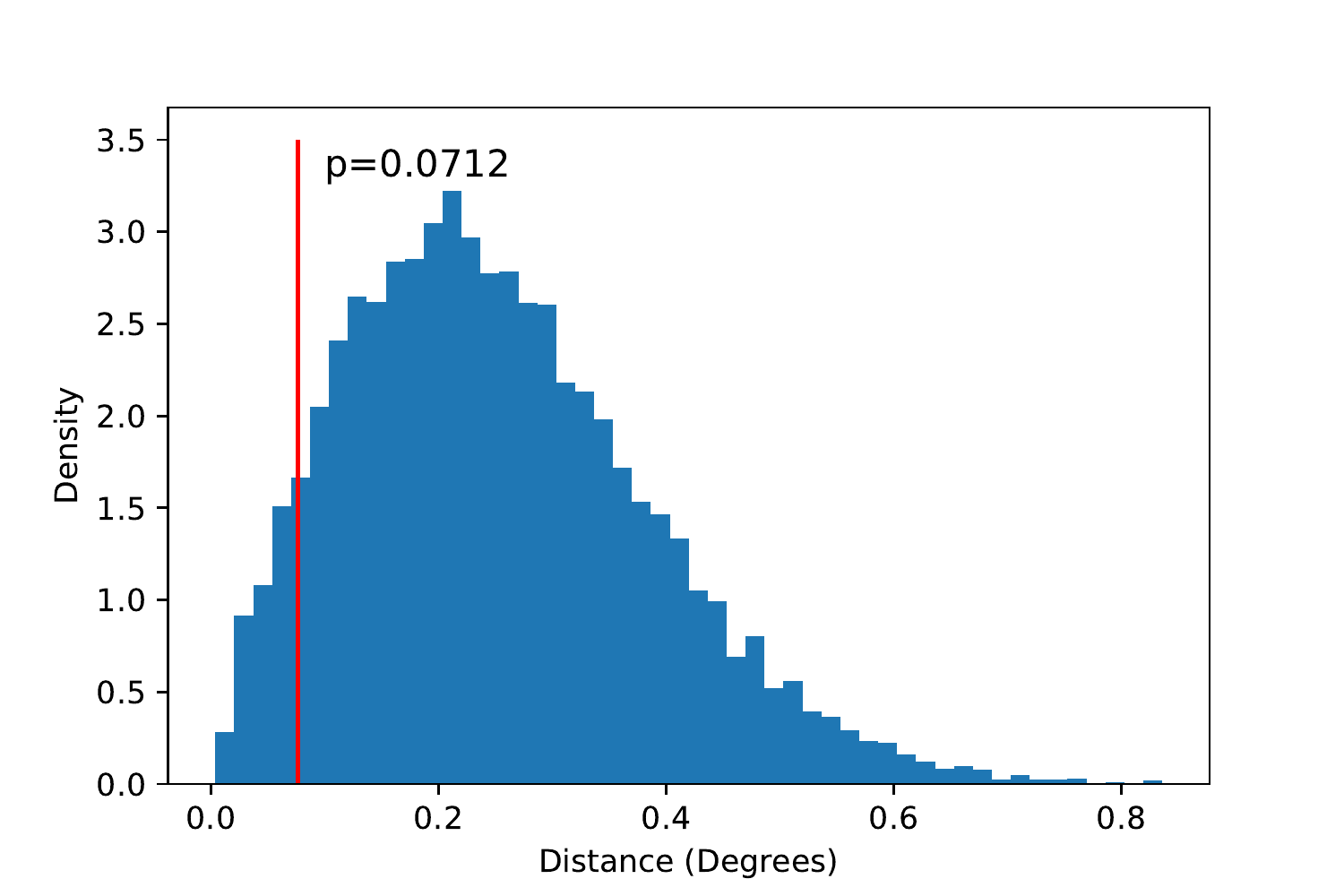}
	\caption{The distribution of the smallest distances between blazars and neutrino events in each simulation. The vertical line denotes the observed value of \ang{0.0763}.}
	\label{fig:bzcat_distance}
	\end{figure}

	The angular distance between TXS 0506+056 and IceCube-170922A is relatively small (\ang{0.0763}). We conduct a total of $10 000$ simulations and record the smallest distance between simulated blazar directions and neutrino alert directions and compare it with the distance between TXS 0506+056 and IceCube-170922A. As shown in Fig. \ref{fig:bzcat_distance}, the probability for the simulation to generate a smaller distance than observed is $0.0712$, or the confidence level for the association between TXS 0506+056 with any neutrino alert based on distance is $1.47 \sigma$. This confidence level is lower than that reported in Ref. \cite{icecube2018multimessenger}.

	\section{Coincidence search of neutrino events in blazar directions}
	\label{sec:Neutrinoinblazar}
	The second approach is to search for neutrino events around the direction of blazars and calculate likelihood using blazar and neutrino event directions.
	
	For this search, we use all the blazar locations from ROMA-BzCat \cite{roma5th}, and all the neutrino events with energies higher than 100 TeV from the all-sky point-source IceCube data in years 2010--2012 dataset \cite{IceCubeCollaboration2018}. This IceCube database includes neutrino events only (F. Halzen, 2020, private communication). This amounts to 48 046 neutrino events. We choose a threshold energy of 100 TeV for two reasons. First, since IceCube-170922A has an energy of 290 TeV, a 100 TeV threshold allows us to select similar events to check their associations with blazars. Second, the neutrino events in our selected catalog show a clear bimodal distribution, with the lower peak dominated by atmospheric neutrinos. 100 TeV is a safe threshold to select neutrinos with a cosmic origin.
	
	To calculate a confidence level of neutrinos clustering around a certain direction produced by an astronomical source, we employ the same unbinned likelihood method for likelihood used by Refs. \cite{turley2018coincidence, krauss2018fermi} as described in Ref. \cite{braun2010time}.
	
	We select all the neutrino events whose angular distances on the celestial sphere to the interested sources are smaller than \ang{20} \footnote{This is about 10 times of the error boxes of the neutrino events. }. Then, for a total of $N$ events, the probability density function for the $i$th event is given by
	$$\frac{n_s}{N}S_i+(1-\frac{n_s}{N})B_i,$$
	where $S_i$ and $B_i$ are the probability density functions (PDFs) for the source and the background event distribution, respectively. $n_s$ is the number of events accredited to the source. The likelihood function of the entire dataset is the product of each individual PDF as a function of $n_s$:
	$$\mathcal{L}(n_s)=\prod_{i=1}^{N}\left[\frac{n_s}{N}S_i+(1-\frac{n_s}{N})B_i\right].$$
	Here, $n_s$ is an unknown variable and should be determined by maximizing the likelihood function. We can then define a test statistic (TS):
	$$TS=2\log\frac{\mathcal{L}(\hat{n}_s)}{\mathcal{L}(0)}.$$
	This is equivalent to the likelihood ratio of two hypotheses:
	$H_1$: Of all the $N$ events, $n_s$ are produced by the interested source.
	$H_0$: All the $N$ events are background events.
	
	The likelihood ratio will indicate how likely it is that some of the neutrino events can be attributed to the source we are trying to test the association with the diffuse neutrino flux.
	
	The source PDF is modeled by the probability for a source to produce a neutrino event the same as a particular event in the data. For this case, this PDF is equal to the point spread function (PSF) of the source. We assume that the PSF for the source neutrino events has a Gaussian distribution, with the mean at the position of the interested source, and the standard deviation $\sigma$ being the reported localization error given in the IceCube data, i.e.
	$$S_i=\frac{1}{2\pi\sigma_i^2}e^{-\frac{\left|r_i-r_s\right|^2}{2\sigma_i^2}} .$$
	We estimate the background PDF by calculating the ratio of average number of events per steradian at each declination and the total number of events. Right ascension is irrelevant here, because of Earth's self-spin. The background PDF for each year in the IceCube dataset is calculated individually, as the equipment setup of IceCube is different in different years.
	
	It is very important to note that according to the definition of PDF, when integrated along the entire parameter space, both the source and background PDFs should yield unity. Therefore, normalization will be necessary; otherwise, the source and background PDFs will have different weights in the likelihood function. The $\frac{1}{2\pi\sigma_i^2}$ normalization for the source PDF and division by the total number of events for the background PDF are needed for this reason.
	
	It is also common in some studies to include an energy spectrum component in PDFs to distinguish cosmic neutrino events from atmospheric events. However, the background term in the unbinned method for likelihood denotes not only the atmospheric background, but also the cosmic background, i.e. neutrino events originated from sources other than the interested source. By introducing the energy spectrum component, we make the assumption that the interested source is the only neutrino source in the studied area; thus, all the neutrinos which do not come from the interested source are atmospheric neutrinos. Given the large localization error of the neutrino events by IceCube and the existence of the isotropic diffuse high-energy cosmic neutrino flux, this assumption is inappropriate. 
	
	Therefore, in this study, we do not include energy spectrum components for the source or background PDFs. We use only events that have high probabilities of being cosmic.
	
	While the likelihood ratio obtained with this method provides a relative measurement of how likely it is to reject the null hypothesis that all the observed neutrino events around the interested source are not produced by that source, it fails to give a straightforward indication of the level of significance. Furthermore, how the fitted number of source events $n_s$ contribute to the association of neutrino events with the interested source is not well defined. If one source has more fitted source events but a lower likelihood ratio, and another source has fewer fitted source events but a higher likelihood ratio, it is difficult to say which of the two sources has the stronger association.
	
	For this reason, for each source with $n_s$ and TS greater than $0$, we conduct a Monte Carlo simulation. We randomly generate neutrino events isotropically distributed around the interested source within a distance of \ang{20} \footnote{Notice that even though the IceCube effective area varies with energy, its dependence on zenith angle is small for the same energy. Also as long as the distribution of the effective area does not rely on the distance to the central source, our results would not be affected.} We then assign energies to each simulated event using the kernel density estimation of the observed neutrino energy distribution of each year. The localization error of each event is subsequently determined from its simulated energy from the relationship of observed event energy and angular resolution given in the IceCube data. The same event selection and unbinned likelihood analysis is carried out, and we can get a distribution of $n_s$ and TS from this Monte Carlo simulation. We can then infer the significance of observed result by comparing it with the distribution and obtain a $p$ value by calculating the percentile observed values lie in.
	
\begin{figure}[ht]
		\centering
		\includegraphics[width=0.5\textwidth]{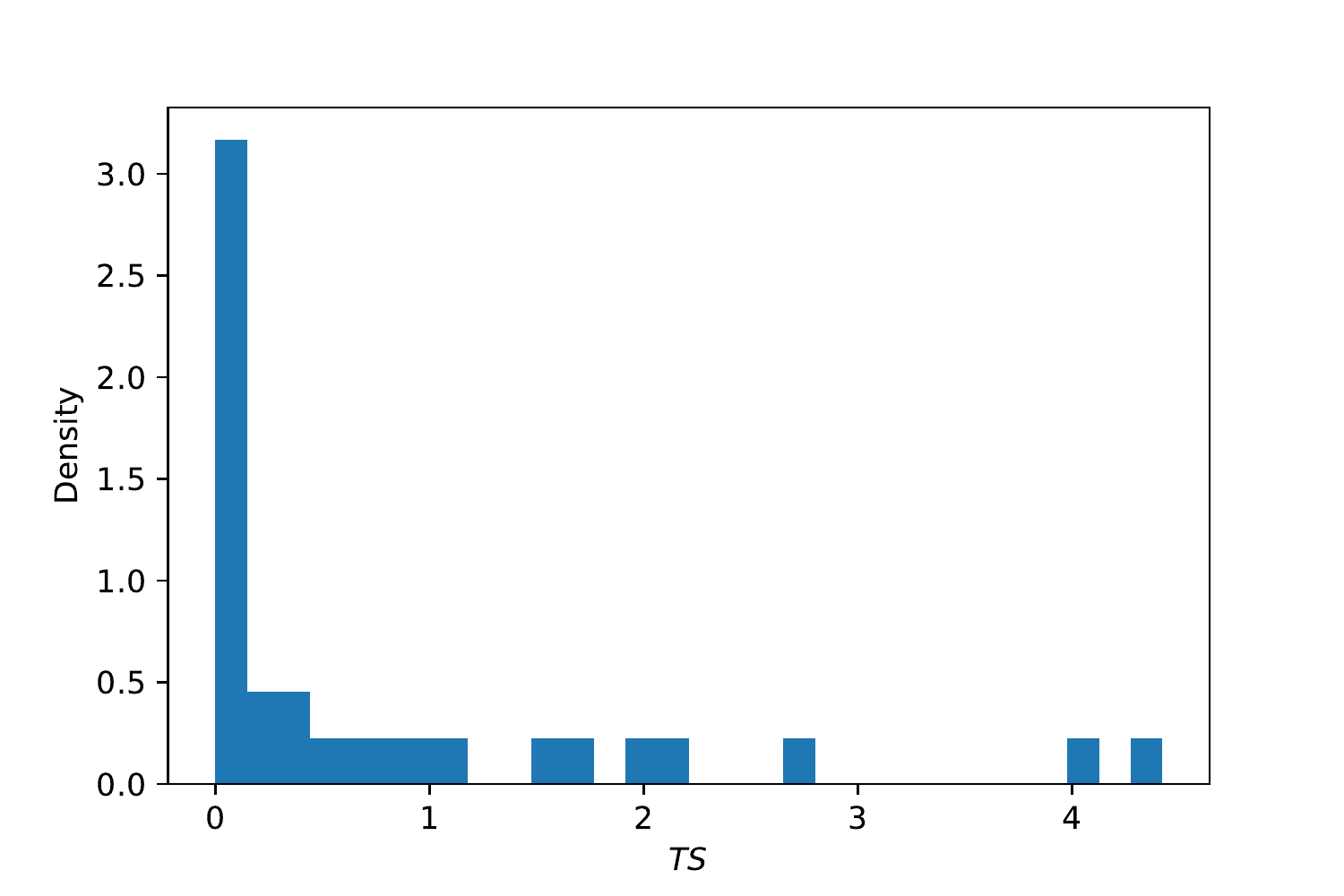}
		\caption{The distribution of the association TS for blazars.}
		\label{fig:TS}
	\end{figure}

	We use the unbinned method to calculate the TS and corresponding confidence level from Monte Carlo simulations for the association of neutrino events and blazars. The distribution of the calculated TS is presented in Fig. ~\ref{fig:TS}. A total of 30 blazars display $n_s$ and TS greater than $0$. The distribution of TS does not deviate too much from normal distribution, as one would expect from a random and isotropic distribution of neutrino events.

%	\begin{table}[]
%		\caption{Blazar sources with test statistics larger than $0$}
%		\begin{ruledtabular}
%			\begin{tabular}{lllllll}
%				Source Name & RA(J2000.0) & DEC(J2000.0) & $\widehat{n_s}$ & TS             & $p_{n_s}$ & $p_{TS}$ \\
%				\hline
%                [VV2006] J224046.9+132602     & 340.195     & 13.434       & 9.0         & 0.293          & 0.089     & 0.107    \\
%			\end{tabular}
%		\end{ruledtabular}
%		\label{tab:blazartable}
%	\end{table}
	
\begin{figure}[ht]
	\includegraphics[width=0.49\textwidth]{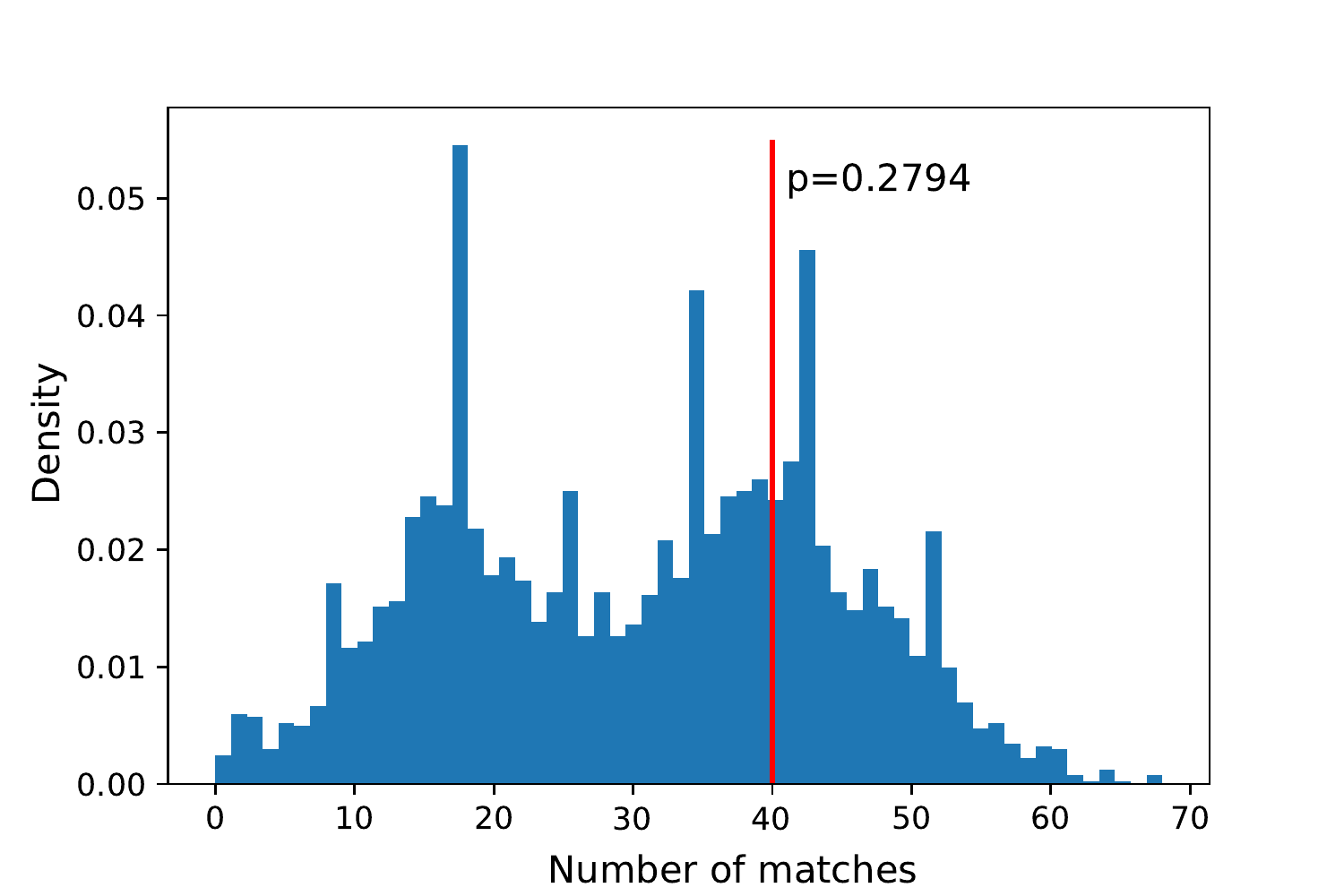}
	\caption{The distribution of the number of matches between all blazars and neutrino events with energies greater than $100$ TeV. The vertical line denotes the number of blazar matches with TXS 0506+056, $40$.}
	\label{fig:bzcat_match}
	\end{figure}

	We again scrutinize the possibility of TXS 0506+056 being a strong neutrino emitter by counting the number of neutrinos whose error region contains the locations of blazars, and compare the numbers with that of TXS 0506+056. The results are shown in Fig. \ref{fig:bzcat_match}. The number of matches for TXS 0506+056 does not deviate much from average.
	
	\section{Coincidence search of neutrino events in directions of Fermi $\gamma$-ray sources}
	\label{Fermi}
	
	The proposed association between TXS 0506+056 and IceCube-170922A is justified by the flaring state of TXS 0506+056 at the same time of the high-energy neutrino event, as it is possible that neutrino fluxes scale with $\gamma$-ray fluxes of astrophysical objects. This is expected within the hadronic model of $\gamma$-ray blazars \cite{2018PhT....71j..36M}.  Indeed, according to the Supplementary Material of original discovery paper \cite{icecube2018multimessenger}, the inclusion of the Fermi LAT data allows the team to reduce the $p$ value to claim an association. On the other hand, IceCube-170922A cannot be accounted for within the simple one-zone model of blazar jets for the parameters of TXS 0506+056 \cite{gao2019modelling,halzen2019neutrino,padovani2018dissecting,sahakyan2018lepto,murase2018blazar,petropoulou2019neutrinos}. Nonetheless, we perform a search for coincident neutrino events from all the monitored steady or flaring Fermi $\gamma$-ray sources, which include both Galactic and extragalactic sources.
	
	We obtain all the light curves of $\gamma$-ray sources monitored by Fermi with weekly bins and set a flux limit of $5\times10^{-7}\ {\rm ph\ cm^{-2}\ s^{-1}}$. This flux limit is the $\gamma$-ray flux level of TXS 0506+056 during its 2017 neutrino flare \cite{padovani2018dissecting}. Then, we apply a time filter to the 3-year IceCube neutrino data, selecting only neutrino events observed during the time when the $\gamma$-ray flux of the interested source is above this flux limit. Likelihoods are subsequently calculated using only selected events around the Fermi source without the use of the temporal information.
	
\begin{figure}[ht]
	\includegraphics[width=0.49\textwidth]{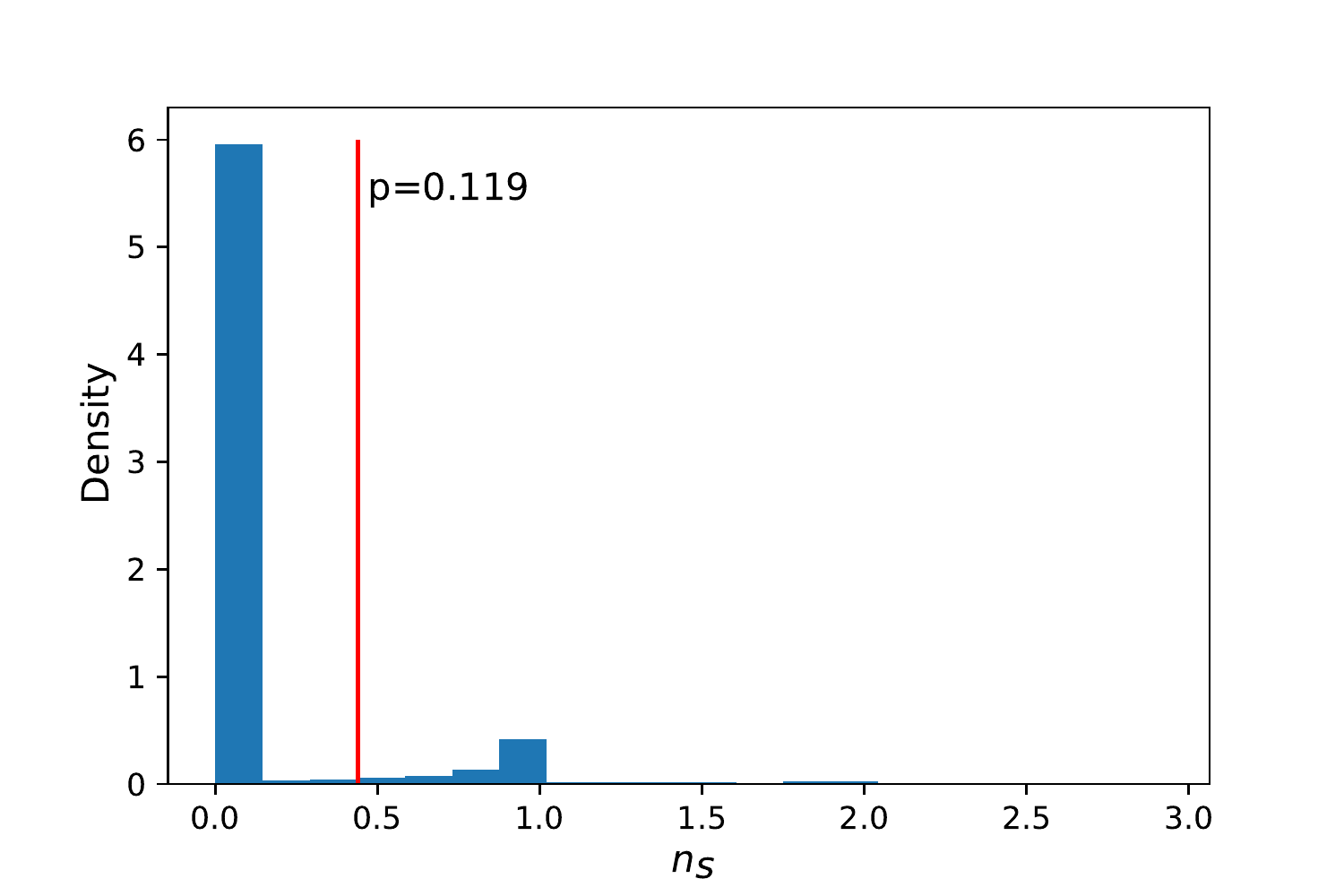}
	\includegraphics[width=0.49\textwidth]{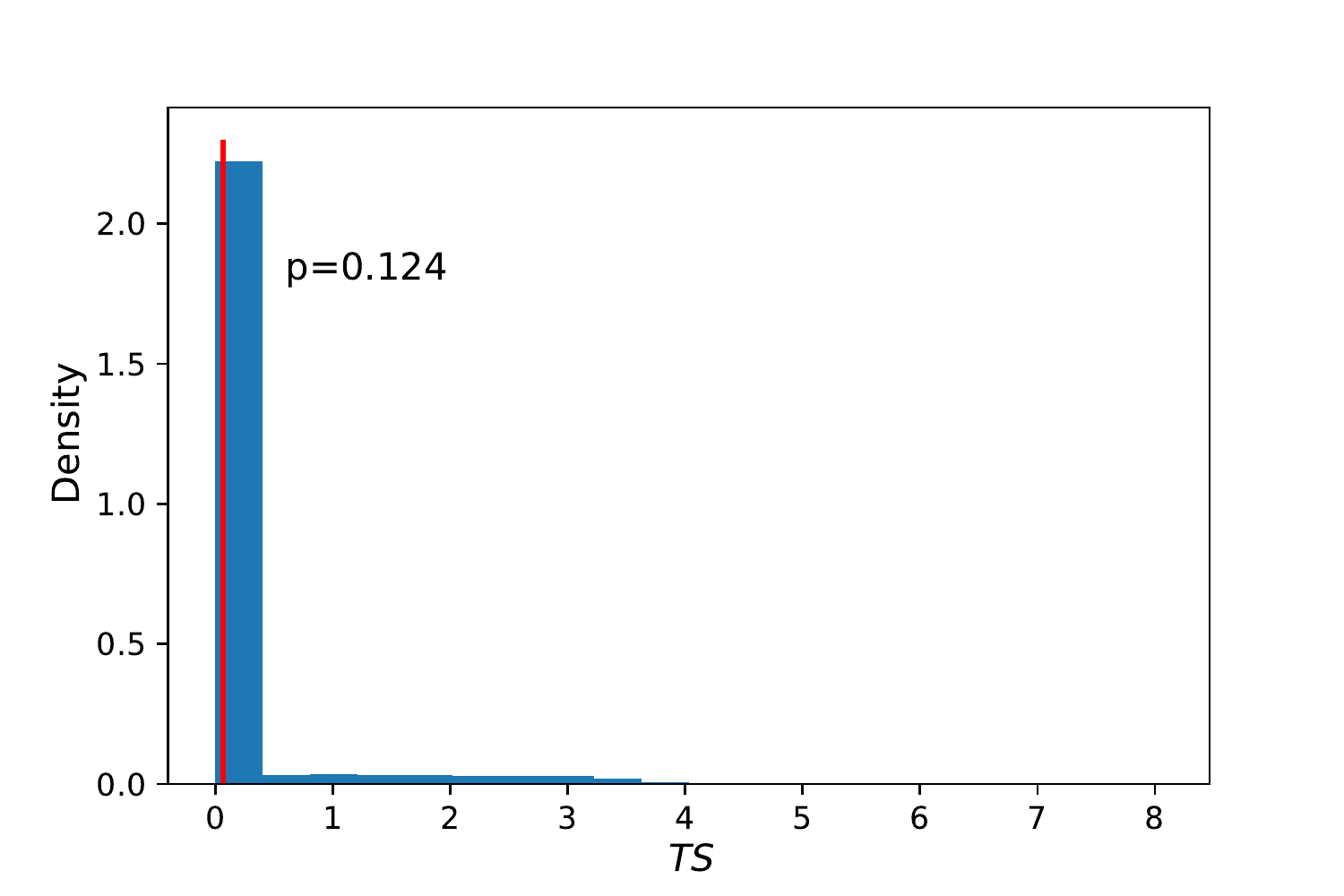}
	\caption{Distributions of $n_s$ and TS from Monte-Carlo simulations for 3C 279. The vertical lines show the observed values of $0.439$ and $0.067$.}
	\label{fig:3c273}
	\end{figure}

	Of all the 170 monitored $\gamma$-ray sources, only 3C 279 displays a TS larger than $0$. However, further Monte Carlo simulations rule out that any of the sources are associated with the diffuse high-energy cosmic neutrino flux, with $p_{n_s}=0.128$ and $p_{\rm TS}=0.134$ from Monte Carlo simulations as shown in Figure \ref{fig:3c273}.
	
%	\begin{table}[]
%		\caption{Fermi sources with test statistics larger than $0$}
%		\begin{ruledtabular}
%			\begin{tabular}{lllllll}
%				Source Name & RA(J2000.0) & DEC(J2000.0) & $\widehat{n_s}$ & TS             & $p_{n_s}$ & $p_{TS}$ \\
%				\hline
%				QSO B2320-035    & 350.883     & -3.285        & 0.942         & 1.877          & 0.109      & 0.096    \\
%			\end{tabular}
%		\end{ruledtabular}
%		\label{tab:fermi}
%	\end{table}

	\section{Conclusions}
	\label{sec:Conclusions}
	In this paper, we reinvestigated the chance probability of the association between TXS 0506+056 and IceCube-170922A utilizing the latest public IceCube data and the Roma-BzCat catalog. We draw the following conclusions:
	\begin{itemize}
		\item
		We calculate the probability of any coincidence between neutrino alerts and blazars directions by simulating blazar directions. We also compare the angular distance between TXS 0506+056 and IceCube-170922A with simulated angular distance with IceCube-170922A or all neutrino alerts. The calculated confidence level for association with any neutrino alerts is only $1.47 \sigma$. Thus, an association between blazars and neutrino alerts, in general, is not supported, even though we cannot rule out the individual IceCube-170922A / TXS 0506+056 association.
		\item
		We conduct a search for clustering of neutrino events around known blazar directions using the unbinned likelihood method. While some source directions show test statistics greater than 0, Monte Carlo simulations with randomly generated neutrino events do not support associations. We also select neutrino events based on the $\gamma$-ray flux of the interested source at each time bin and apply the same method. Again, no significant association was found.
		\item
		While current models for neutrino generation in blazars predict that neutrinos will appear alongside $\gamma$-ray photons, there are many other processes that can generate $\gamma$-ray photons. Thus, a spike in the $\gamma$-ray flux of a source does not guarantee that the neutrino flux from the same source will also increase significantly. In this study, we show that, at the reported $\gamma$-ray flare fluxes, the majority of sources do not show a corresponding neutrino flare. Thus, it is difficult to use $\gamma$-ray flux of a source to support the association of that source with neutrinos. 
	\end{itemize}
	
	Based on the above analysis, we can conclude that if the IceCube-170922A / TXS 0506+056 association is real, it must be a peculiar case, and some special conditions are needed to make a blazar become a bright high-energy neutrino source. This point was recently also emphasized by Refs. \cite{halzen2019neutrino,halzen19b}. Indeed, a new analysis on VLBA observations of TXS 0506+056 revealed that this blazar may have a structure of two jets during a collision course, which may explain the excess neutrino flux compared with other blazars \cite{britzen2019cosmic}. The recent discovery of another possible association between FSRQ PKS 1502+106 and IceCube-190730A may also fall into the similar category \cite{lipunov2019icecube}. However, the calculation presented in Ref. \cite{britzen2019cosmic} (in particular, Eq.(14) in that paper) overestimated the $p\gamma$ optical depth by $\sim 10$ orders of magnitude \footnote{We thank Ruoyu Liu for pointing out this calculation error.}. Further theoretical modeling is desirable to see whether interacting galaxies with radio jets can indeed facilitate the production of high-energy neutrinos, making them credible sources of the neutrinos seen by IceCube.

	\begin{acknowledgments}
        The authors thank Francis Halzen for helpful information and constructive comments. This work is supported by the Top Tier Doctoral Graduate Research Assistantship (TTDGRA) at University of Nevada, Las Vegas.
    \end{acknowledgments}

	\bibliography{document}
\end{document}